\documentclass[twocolumn]{aastex62}

\usepackage{amsmath}

\defcitealias{Yuan:18b}{Yuan18}

\shorttitle{The role of hot AGN feedback mode}
\begin{document}

\title{On the Role of Hot Feedback Mode in Active Galactic Nuclei Feedback in an Elliptical Galaxy}

\email{d.yoon@uva.nl; fyuan@shao.ac.cn}

\author[0000-0001-8694-8166]{Doosoo Yoon}
\affiliation{Key Laboratory for Research in Galaxies and Cosmology, Shanghai Astronomical Observatory,
Chinese Academy of Sciences, 80 Nandan Road, Shanghai 200030, China}
\affiliation{Anton Pannekoek Institute for Astronomy, University of Amsterdam,
Science Park 904, 1098 XH Amsterdam, the Netherlands}

\author[0000-0003-3564-6437]{Feng Yuan}
\affiliation{Key Laboratory for Research in Galaxies and Cosmology, Shanghai Astronomical Observatory,
Chinese Academy of Sciences, 80 Nandan Road, Shanghai 200030, China}
\affiliation{School of Astronomy and Space Science, University of Chinese Academy of Sciences, 100049, Beijing, China}

\author[0000-0002-6405-9904]{Jeremiah P. Ostriker}
\affiliation{Department of Astronomy, Columbia University, 550 W. 120th Street, New York, NY 10027, USA}
\affiliation{Department of Astrophysical Sciences, Princeton University, Princeton, NJ 08544, USA}

\author[0000-0002-5708-5274]{Luca Ciotti}
\affiliation{Department of Physics and Astronomy, University of Bologna, via Piero Gobetti 93/2, 40129 Bologna, Italy}

\author[0000-0003-3564-6437]{Bocheng Zhu}
\affiliation{Key Laboratory for Research in Galaxies and Cosmology, Shanghai Astronomical Observatory,
Chinese Academy of Sciences, 80 Nandan Road, Shanghai 200030, China}

\begin{abstract}
Depending on the value of the accretion rate, black hole accretion is divided into cold and hot
modes. The two modes have distinctly different physics and correspond to two feedback modes. Most
previous feedback works  either focus only on one mode, or the accretion physics is not always
properly adopted. Here, by performing hydrodynamical numerical simulations of AGN feedback in an
elliptical galaxy, we show that including both is important, and gives different results from
including just one or the other. We specifically focus on the wind and radiation (but neglecting the
jet) feedback in the hot mode, to explore whether this particular mode of feedback can play any
important feedback role.  For this aim, we have run two test models. In one model, we always adopt
the cold mode no matter what value the accretion rate is; in another model, we turn off the AGN once
it enters into the hot mode. We have calculated the AGN light curves, black hole growth, star
formation, and AGN energy duty-cycle; and compared the results  with the model in which the two
modes are correctly included.  Important differences are found. For example, if we were to adopt
only the cold mode, the total mass of newly formed stars would become two orders of magnitude
smaller, and the fraction of energy ejected within the high accretion regime (i.e., $L_{\rm BH} >
2\% L_{\rm Edd}$) would be too small to be consistent with observations.  We have also investigated
the respective roles of wind and radiation in the hot mode.
\end{abstract}

~~~~~

\section{Introduction}

More and more evidence has shown that a supermassive black hole (SMBH) may evolve together with its
host galaxy, such as the strong correlation between the mass of the black hole and the luminosity,
stellar velocity dispersion, or stellar mass in the galaxy spheroid (\citealt{Magorrian:98};
\citealt{Tremaine:02}; see review by \citealt{Kormendy:13}). Both observational and theoretical
studies now strongly indicate that such a co-evolution between the black hole and the galaxy is
the consequence of active galactic nuclei feedback \citep{DiMatteo:05, Springel:05, Murray:05,
Croton:06, Ciotti:07, Ciotti:09a, Sijacki:07, Hopkins:08, Ostriker:10, Fabian:12, Hirschman:14,
King:15, Naab:17, Weinberger:18}. In this process, the outputs from the AGN, namely radiation,
wind, and jet, interact with the interstellar medium (ISM) in the host galaxy and alter the
density and temperature. Consequently, star formation activity is changed, and thus galaxy
evolution is affected. The change of the properties of ISM will in turn affect the fueling of
the AGN and the evolution of the black hole mass.  In this scenario, the value of AGN accretion
rate and the model of accretion physics are of central importance for evaluating the effects
of AGN feedback since they determine the magnitude of AGN outputs.

Black hole accretion comes in two different modes, namely cold and hot modes, depending on the
value of the mass accretion rate of the central black hole. The boundary between the two modes
is $\sim 2\% L_{\rm Edd}$ \citep{Yuan:14}.  Simulations of AGN feedback have shown that the
black hole activity usually oscillates with time, passing through both modes. The cold and hot
accretion modes correspond to the cold and hot feedback modes\footnote{In the literature, the
cold mode is also called ``quasar'' or ``radiative'' mode, while the hot mode is often called
``radio'' or ``jet'' or ``kinetic'' or ``maintenance'' mode.}.  In this paper we will focus
on the role of hot feedback mode.  Many previous studies  have taken into account  this mode
\citep[e.g.,][]{Croton:06,Bower:06,Sijacki:07,Ostriker:10,Mcnamara:12,Li:14,Li:15,Schaye:15,
Guo:16,McAlpine:17, Weinberger:17, Weinberger:18, Habouzit:18,Guo:18}. For example, by using
semi-analytical approach, \citet{Croton:06} found that this mode of feedback can provide an
efficient source of energy to solve the cooling flow problem,  the exponential cutoff at the bright
end of the galaxy luminosity function, and the increased mean stellar age in massive elliptical
galaxies. In the hydrodynamical simulation work of \citet{Sijacki:07}, they included both the
cold and hot modes in their ``sub-grid'' model and applied it to galaxy cluster formation. In
a more recent work, by invoking winds launched in the hot mode, \citet{Weinberger:17} found
that their cosmological simulation model can overcome some serious problems in previous galaxy
formation models and successfully produce red, non-star-forming massive elliptical galaxies,
and achieve realistic gas fraction, black hole growth histories and thermodynamic profiles
in large haloes.  This feedback model was later applied to the cosmological simulation of
Illutris-TNG \citep{Weinberger:18}.

Cosmological AGN feedback works  focus on very large scales, so they can simulate the formation
and evolution of numerous galaxies. But their resolution has to be rather low, e.g., several
$\rm {kpc}$; it is thus difficult to study the details of AGN feedback. Another approach is
to focus on the much smaller  galaxy scale \citep{Ciotti:97, Ciotti:01, Ciotti:07, Novak:11,
Novak:12, Gaspari:13a, Gan:14, Hopkins:16, Ciotti:17b, Yuan:18b, Yoon:18, Li:18}. The advantage
of this kind of model is that we can easily resolve the boundary of black hole accretion flow,
i.e., the Bondi radius, which is typically several tens of $\rm pc$. In this case, the accretion
rate can be reliably calculated, so we can avoid the uncertainty as large as $\sim 100$  or even
larger in  cosmological simulations \citep{Negri:17,[][see also]Ciotti:17a,Ciotti:18}. Moreover,
we can carefully investigate the details of the AGN feedback, i.e., how the AGN outputs interact
with the ISM.  This is the approach adopted in the present work.

An equally important issue for the AGN feedback study, in addition to the precise determination
of the mass accretion rate, is the AGN physics, which describes the output from the AGN for a
given accretion rate. After several decades' observational and theoretical efforts, we now have
accumulated quite solid knowledge of black hole accretion \citep[see reviews by][]{Pringle:81,
Blaes:14, Yuan:14}.  Based on these knowledge, \citet[hereafter Yuan18]{Yuan:18b} has presented a
model framework of AGN feedback study by incorporating state-of-art AGN physics. This framework
has been adopted in several later works of a series  \citep{Yoon:18, Li:18, Gan:19} and these
works focus on developing different aspects of the model:  \citet{Yoon:18} has extended the
low-angular momentum galaxy in \citetalias{Yuan:18b} to the case of high angular momentum;
\citet{Li:18} has compared the different roles played by AGN feedback and various components
of stellar feedback (supernovae and stellar wind); while \citet{Gan:19} has focused on the role
of gravitational instability of the galactic circum-nuclear, cold gas disk.

One crucial point of the accretion physics is that the radiation and wind outputs from the AGN
are distinctly different in the two modes and they cannot be described by the same scaling.
Roughly speaking, both the radiation and wind outputs would be significantly over-estimated
in the hot mode if we were to choose the scaling of the cold mode.  However, most current AGN
feedback work focuses on either cold mode or hot mode. Even if they include both, the accretion
physics, especially the physics of the hot mode, is not correctly adopted, since they often
adopt the same scaling for the wind and radiation outputs as the cold mode, which is incorrect.

In this paper, we show how important it is to include both modes by performing high-resolution
hydrodynamical simulations following \citetalias{Yuan:18b}. We specifically focus on the roles
of wind and radiation feedback in the hot mode\footnote{We neglect the jet in our model, we will
discuss this issue further in section 2.4.}. Both observational and theoretical simulations have
shown that the AGN reside in the hot mode for a much longer time than in the cold mode (e.g.,
\citealp{Haiman:01, Heckman:04, Greene:07, Kauffmann:09, Shankar:10}; \citetalias{Yuan:18b};
\citealp{Yoon:18, Gan:19}. For instance, a statistical study has shown that the percentage of
the time spent in the active phase of the black hole, which roughly corresponds to the cold
accretion mode, is reported to be $\sim$0.4\% or even smaller \citep{Greene:07}. Therefore,
although the wind and radiation in the hot mode are much weaker than those in the cold mode,
it is hard to exclude the possibility that the hot mode feedback still plays an important role
due to its cumulative effects.

This paper is structured as follows. In \S\,\ref{sec:models}, we  overview different components
of our model, including the simulation setup, the galaxy model we adopt, the calculations of
star formation rate, the stellar feedback, and the physics of cold and hot accretion modes.
In \S\,\ref{sec:results}, we describe our simulation results.  We summarize our results in
\S\,\ref{sec:summary}.

~~~~

\section{Model}\label{sec:models}

The code that this work will be based on is called ``{\it MACER}'', which has been
described in \citetalias{Yuan:18b} (see also \citealt{Yuan:18a} for a short description
of the main components of the model) and is based on the code originally developed by
\citet{Ciotti:97,Ciotti:01,Ciotti:07} over several years. Specifically, using this code,
we study the AGN feedback in an isolated elliptical galaxy.  The Bondi radius is resolved
and thus the accretion rate can be calculated precisely. We then calculate the AGN outputs
including the wind and radiation based on the sub-grid accretion physics; the interaction of
radiation and wind with the interstellar medium is also calculated by considering simplified
radiative transfer. Other physical processes such as star formation and energy release through
stellar evolution, including supernova (SNe) I and II, are taken into account in a way we will
describe below. Importantly, we also include the very significant mass input to the ISM from
planetary nebulae and AGB stars.  For the convenience of the readers, in the following sections
we summarize the key components of the model.

\subsection{Simulation Setup}

Simulations were performed with the parallel ZEUS-MP/2 code \citep{Hayes:06} using axisymmetric
spherical coordinates ($r,\theta$). The black hole is located at the origin. We adopt a
logarithmic mesh with 120 grids in the radial direction, which covers the range of 2.5 pc --
250 kpc. We use the standard ``outflow boundary condition'' in the inner and outer radial
boundary. In $\theta$ direction,  30 grids are uniformly divided. The highest resolution
is thus achieved at the innermost region, which is $\sim 0.3$ pc. Such a configuration is
essential, since the innermost region is the place where the radiation and the wind originate
and where the accretion rate of the black hole is determined, and thus are most important for
AGN feedback. Even for the hot gas in the galactic center region, the Bondi radius is $\sim 6$
pc; for the cold gas, the Bondi radius is much larger \citepalias{Yuan:18b}. So  the  inner
boundary of our simulation domain is at least two times smaller than the Bondi radius. This
means that we can precisely calculate rather than estimate the black hole accretion rate,
which is crucial to evaluate the effects of the feedback.

Within the inner boundary, the accretion flow still cannot be resolved. We treat the accretion
process in that region as sub-grid physics. We describe this part in \S~\ref{subsec:agn_model}
\citepalias[see also][]{Yuan:18b}.

\subsection{Galaxy Model}

In this work, we focus on the secular evolution of an isolated elliptical galaxy. We only
consider the gas produced by the stellar evolution as the material for black hole accretion. 
In reality, there must be some external gas supply to the galaxy. One is the cosmological inflow
from intergalactic medium. In addition, for those galaxies located at the center of galaxy
clusters or galaxy groups, there will be enormous potential supply of gas from their gaseous
halo, the infall of which causes the ``cooling flow'' problem. Because of these limitations,
our present work provides a necessary (but not sufficient) condition for AGN feedback to be
important for the quenching of galaxies. And last, mergers are also neglected. We make
this assumption because the majority of nuclear activity in the universe has taken place due
to internal dynamical instabilities rather than from violent mergers \citep{Cisternas:11}.
Observations indicate that, at least for moderate-luminosity AGNs, the growth of the black
hole and star-forming galaxies have been regulated dominantly by internal secular processes
\citep{Kocevski:12,Fan:14}. But galaxy merger processes may still be effective in fueling gas
to the central black hole \citep{Mihos:96,DiMatteo:05}.   We will study the  effect of including
external gas supply and galaxy merger in our future work.

In our simulation, we start with very low gas density in the galaxy  as an initial condition.
Over a cosmological time, the total mass loss from the passively evolving stars in elliptical
galaxies can reach up to $\sim$20\% of the initial stellar mass \citep{Ciotti:12}, mainly by red
giant winds and planetary nebulae. This is  two orders of magnitude larger than the mass of the
initial central black hole. If all of this gas was accreted into the black hole,  the mass of the
hole would be $\sim$100 times larger than what is observed, $M_{\rm BH}\simeq 10^{-3}\,M_{\star}$
\citep[e.g.,][]{Magorrian:98,Kormendy:13}. Again, we leave the study of including additional initial
gas in the galaxy to a future work.

Following \citet{Ciotti:09b}, we adopt the \citet{Jaffe:83} profile for the initial stellar distribution,
\begin{equation}
    \rho_{\star} = \frac{M_{\star}\,r_{\star}}{4\pi\,r^2 (r_{\star}+r)^2},
\end{equation}
where $M_{\star}$ is the total stellar mass and is set to be $M_{\star}=3\times 10^{11}\,
M_{\odot}$, $r_{\star}$=6.9 kpc is the scale length of the galaxy, which corresponds to the
projected half-mass radius of $R_{e}$=5.14 kpc. The migration of stars is not taken into account
in our simulation; instead, both the initial stars and the newly born stars keep their locations
all of the time.

Along with the central black hole, the dark matter halo and a stellar spheroid are considered as the
dominant contributors of the gravitational potential in the galaxy. The self-gravity of the ISM is
ignored in our simulation. The density profile of the dark matter halo is assumed to be spherically
symmetric. The total mass profile (dark matter plus stars) decreases as $r^{-2}$, as observed
\citep[e.g.,][]{Rusin:05,Czoske:08,Dye:08}.  To simplify our problem, we further assume that the
stars rotate slowly. Since the gas in our simulation comes from the evolving stars, this means that
the specific angular momentum of the gas is small and we do not need to deal with the angular
momentum transfer. For the case of high angular momentum, readers can refer to \citet{Yoon:18,
Gan:19}.

And last, the initial black hole mass is set by the empirical correlation between the black
hole mass and the stellar mass given in \citet{Kormendy:13}, which for the adopted galaxy model
gives $M_{\rm BH,init}=1.8\times 10^9 \, M_{\odot}$.

\subsection{Star Formation and Stellar Feedback}

The cold gas reservoir in the central galaxy is an ideal place for the onset of the radiative
cooling instability, which leads to active star formation. We calculate the star formation rate
per unit volume by means of a recipe that reproduces quite well the standard Schmidt-Kennicutt
empirical relation, as in our previous works \citep[e.g.,][]{Novak:11, Yuan:18b},
\begin{equation}
    \dot{\rho}_{\rm SF} = \frac{\eta_{\rm SF}\,\rho}{\tau_{\rm SF}},
\end{equation}
where $\eta_{\rm SF}$ is the star formation efficiency, $\tau_{\rm SF}$ is the star formation  timescale,
\begin{equation}
\tau_{\rm SF}={\rm max}(\tau_{\rm cool},\tau_{\rm dyn}),
\end{equation}
where the cooling timescale and the dynamical timescale are,
\begin{equation}
\tau_{\rm cool}=\frac{E}{C}, ~~\tau_{\rm dyn}={\rm min}(\tau_{\rm ff},\tau_{\rm rot})
\end{equation}
with
\begin{equation}
    \tau_{\rm ff} = \sqrt{\frac{3\pi}{32G\rho}}, ~~\tau_{\rm rot}=\frac{2\pi r}{v_k(r)}
\end{equation}
where $E$ is the internal energy density, $C$ is the effective radiative cooling rate per unit
volume, $G$ is the Newtonian gravitational constant, $v_k(r)$ is the Keplerian velocity at radius
$r$. The radiative cooling rate $C$ is computed by using the formulae in \citet{Sazonov:05}. It
describes the net heating/cooling rate in photo-ionization equilibrium with a radiation field
corresponding to the average quasar spectral energy distribution.  In particular, line and
continuum heating/cooling, bremsstrahlung losses, and Compton heating/cooling are taken into
account. We note that for simplicity we have ignored the chemical evolution of ISM and the
existence of the dust in the galaxy, which can affect the star formation process.

Different from our previous works \citepalias[e.g.,][]{Yuan:18b}, which only adopted the
above-mentioned standard Schmidt-Kennicutt prescription to calculate the star formation rate,
in the present work we add the following additional constraints. That is, we do not allow star
formation when the gas temperature is higher than $4\times 10^4 {\rm K}$ nor the density is
lower than $1 \,{\rm cm}^{-3}$. This is to mimic the fact that stars are formed from cold
and dense molecular gas. In addition, we now choose a lower value of the star formation
efficiency. \citetalias{Yuan:18b} adopt $\eta_{\rm SF}=0.1$  while $\eta_{\rm SF}=0.01$ is
adopted in the present work.

The calculation of stellar evolution in our simulation follows the description presented in
\citet{Ciotti:12}. Both the stellar winds and SN explosions will provide sources of mass and
energy to the galaxy, and these effects will be taken into account in our simulations. This gas,
when it cools due to radiation, can form stars. Some newly formed massive stars evolve quickly
and explode via SNe II. They will then eject mass and energy into the ISM at some rates, and
this has also been considered in our simulation.

\subsection{Physics of the cold and hot accretion modes}
\label{subsec:agn_model}

\begin{figure*}[!htbp]
    \begin{center}$
        \begin{array}{cc}
            \includegraphics[width=0.48\textwidth]{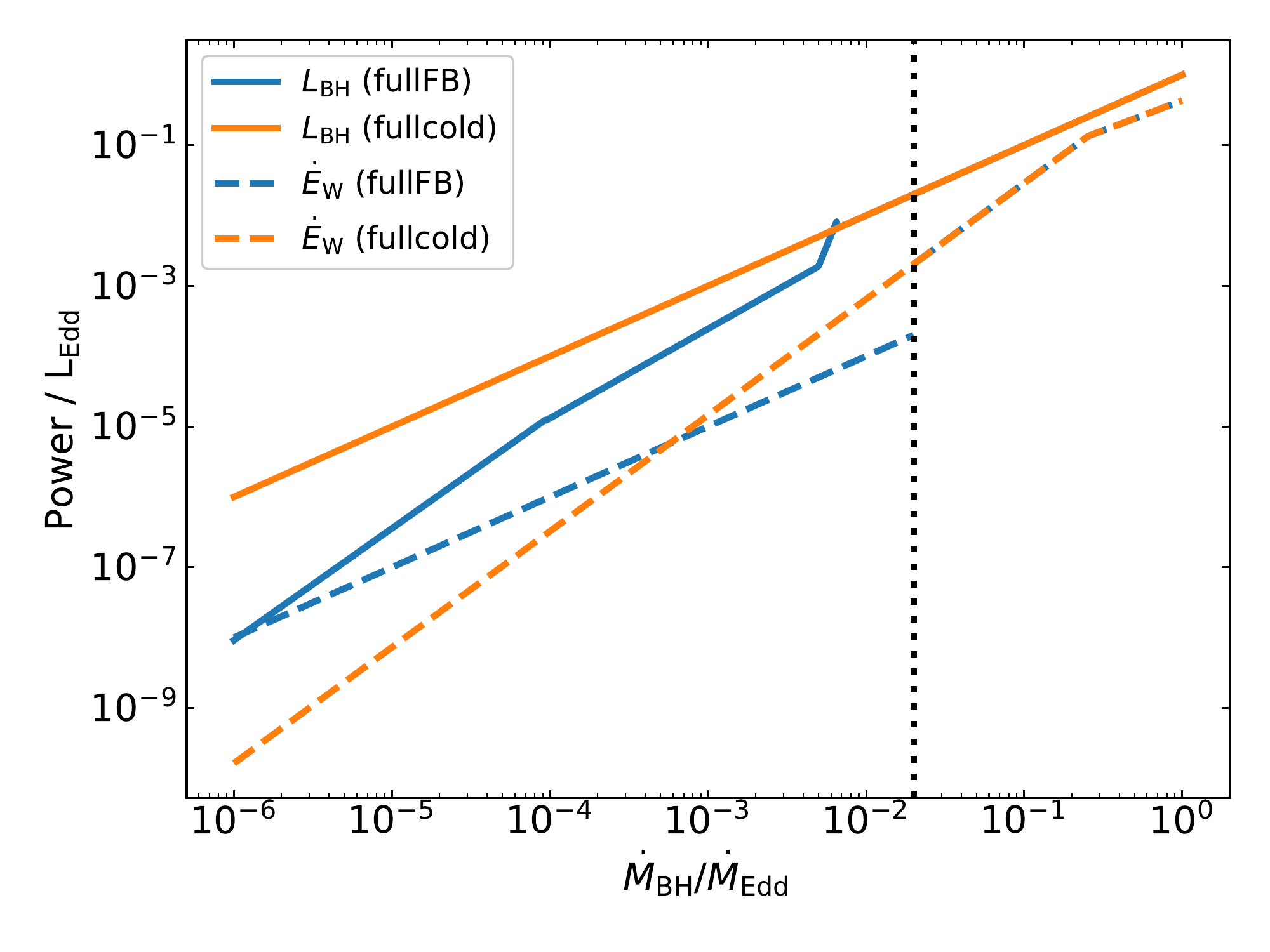} &
            \includegraphics[width=0.48\textwidth]{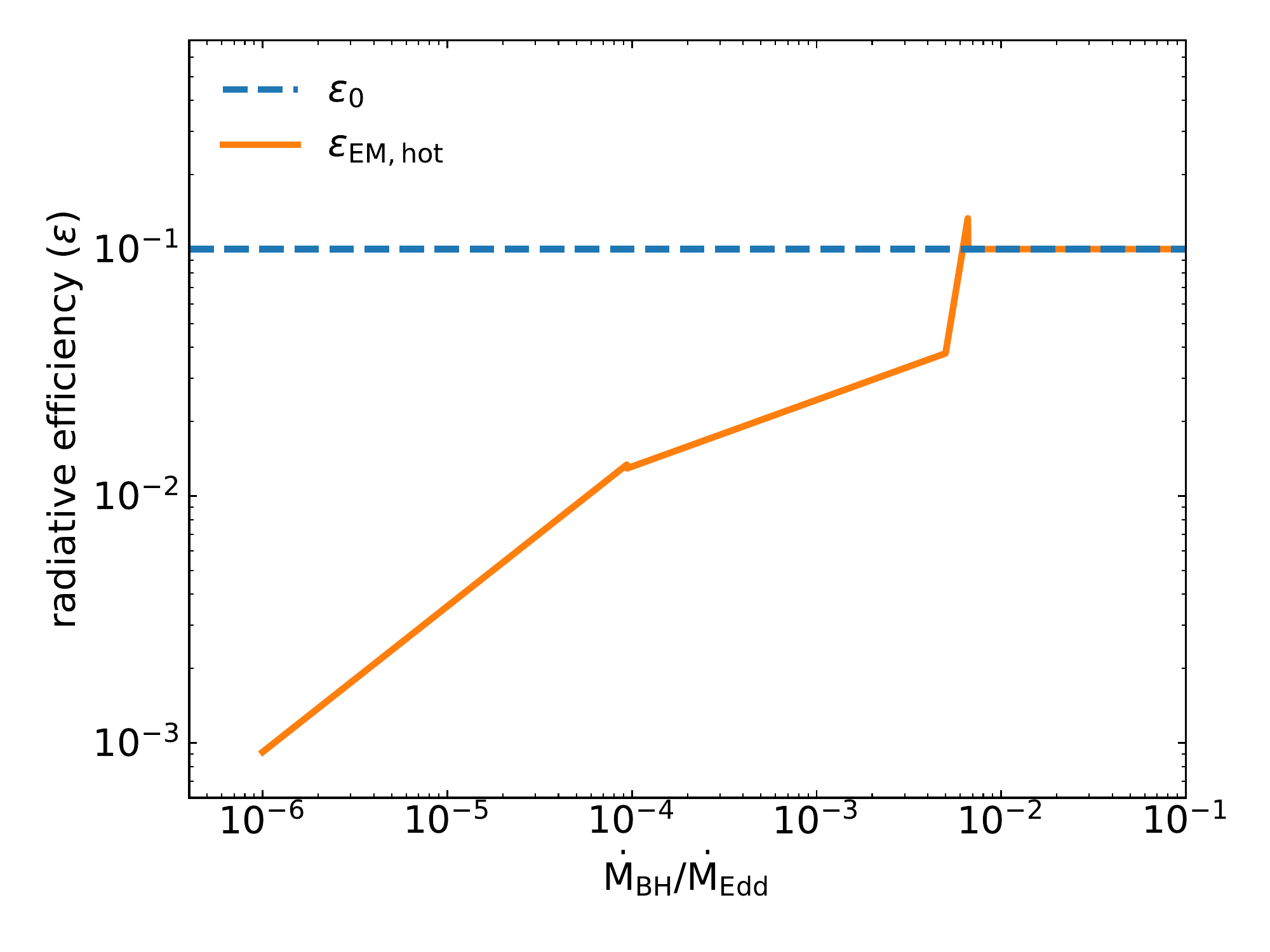}
        \end{array}$
    \end{center}
    \caption{Left: The comparison of AGN luminosity and kinetic power
    as a function of accretion rate between the fullFB model and fullcoldFB model.
    Right: The radiative efficiency for the cold mode ($\epsilon_0$, dashed blue line)
    and the hot mode ($\epsilon_{\rm EM,hot}$, solid orange line) as a function of accretion rate.}
    \label{fig:anal}
\end{figure*}

Depending on the value of the mass accretion rate, black hole accretion has two sets of solutions.
When the accretion rate is higher than $\sim 2\%\dot{M}_{\rm Edd}$, here the Eddington accretion
rate $\dot{M}_{\rm Edd}\equiv 10L_{\rm Edd}/c^2$, the accretion is in the ``cold'' mode, since the
temperature of the accretion flow is relatively low; when the accretion rate is lower than this
value, the accretion will be in the ``hot'' mode, since the temperature of the accretion flow is
several orders of magnitude higher, nearly virial\footnote{In fact, theoretically the cold mode
solution may still exist when the accretion rate is lower than $2\%\dot{M}_{\rm Edd}$.  However,
observations of black hole X-ray binaries show that in this case only the hot mode accretion can
be realized in nature \citep{McClintock:06,Yuan:14}. This is called ``strong ADAF principle'' in
\citet{Narayan:95}.}. The representative solutions in the cold and hot modes are the standard thin
disk solution \citep{Shakura:73} and the advection-dominated accretion flow (ADAF;
\citealt{Narayan:95, Yuan:14}).

The threshold of $2\%\dot{M}_{\rm Edd}$ applies to the accretion rate in the innermost region of the
accretion flow, which is usually smaller than the rate at the inner boundary of the simulation
domain $R_{\rm in}$, partly because of the existence of wind within the accretion flow.  However, in
our model we directly use the threshold of $2\%\dot{M}_{\rm Edd}$ at $R_{\rm in}$ to discriminate
between the two modes. This is because the wind is weak close to or above this accretion rate, as we
will see later in this section.

Since we cannot resolve the scale within $R_{\rm in}$, we have to use some sub-grid physics. When
the accretion is in the cold mode,  the gas will first freely fall until a small accretion disk is
formed with the size of the circularization radius. The accretion rate close to the black hole is
calculated based on this scenario. Readers can refer to section 2.2.1 of \citetalias{Yuan:18b} for
details of the calculation. When the accretion is in the hot mode, the accretion flow will be in the
form of a thin disk at large radii; at a certain radius, $R_{\rm tr}$, the thin disk will be
truncated and transit into a hot accretion flow \citep{Yuan:14}.

Based on the above scenario, once we have obtained the accretion rate at $R_{\rm in}$, we can
calculate the outputs from the AGN based on our knowledge of black hole accretion. The outputs in
general include radiation, wind, and a jet. In the present work, we only consider wind and radiation
but neglect the jet. This is partly because, although jet feedback is generally believed to play an
important role on the galaxy cluster scale \citep[but different opinions still exist, e.g.,][]
{Vernaleo:06, Guo:16, Guo:18}, no consensus has been reached about whether the jet
is important on the galaxy scale. This is because, although jets can be very powerful, since they
are well collimated, they may just pierce through the galaxy without depositing significant energy
in the ISM. In addition, it is still an open question  which one, wind and jet, is more powerful.
Theoretically, in the case of a non-spinning black hole, \citet{Yuan:15} has shown that both the
energy and momentum fluxes of wind are significantly larger than jet\footnote{In the case of a
non-spinning black hole, jet can still be powered by the rotating accretion flow \citep{Yuan:15}.}.
But in the case of a rapidly spinning black hole, the jet may become significantly stronger than
winds due to the additional powering of jet by the rotating black hole (Yang, Yuan et al. 2019, in
preparation). Moreover, it is unlikely that the black hole spin is perfectly aligned with the
angular momentum of the accretion disk, resulting in a precessing jet \citep{Falceta:10}.  Such a
precession may help the interaction between the jet and ISM.  We will investigate the role of jet in
the future.

The dynamics of the accretion flow and wind in the cold and hot modes are very different
\citep{Yuan:14}. For the wind in the cold mode, three mechanisms have been proposed for
the formation of wind, namely thermal, radiation line-force, and magnetic field. But due to
the technical difficulties, theoretical works  works usually focus only on one mechanism,
and so far no consensus has been reached as to the dominant mechanism of wind formation. On
the observational side, however,  we have accumulated abundant observational data on winds
\citep[e.g.,][]{Crenshaw:03, Arav:08, Tombesi:12, King:15}. Following \citetalias{Yuan:18b}, in the
present work  we adopt the statistical results of wind properties obtained in \citet{Gofford:15},
which are obtained by fitting the observations of the wind from 51 AGNs observed by {\it
Suzaku}. The mass flux and velocity of wind are found to be a function of the AGN bolometric
luminosity $L_{\rm bol}$, described by the following equations,
\begin{equation}
    \dot{M}_{\rm W,C} = 0.28 \left( \frac{L_{\rm bol}}{10^{45}\,\rm erg\,s^{-1}} \right)^{0.85} \, M_{\odot}\,\rm yr^{-1},
    \label{coldwindflux}
\end{equation}
\begin{equation}
    v_{W,C} = 2.5\times10^{4}\, \left( \frac{L_{\rm bol}}{10^{45}\,\rm erg\,s^{-1}} \right)^{0.4}\, \rm km\,s^{-1}.
\label{coldwindvelocity}
\end{equation}
We set the largest velocity of the wind to be $10^{5}\, \rm km\,s^{-1}$. But we would like to
emphasize that, a large degree of diversity of wind properties  exists in different observational
results. The effect of wind parameters on AGN feedback will be investigated in our next work
(Yao et al. in preparation).

As for the angular distribution of the mass flux of wind from the cold disk, following our
previous works (e.g., \citetalias{Yuan:18b}), we simply assume that the mass flux of wind
$\propto {\rm cos}(\theta)$, implying that the wind is strongest toward the polar region. This
is in rough accord with the observational prevalence of BAL winds for roughly 1/3 of active
AGN \citep{Arav:08}. We note that since our galaxy model is almost spherically symmetric,
the orientation of the wind is not important.

The bolometric  luminosity from the cold accretion flow is described by
\begin{equation}
    L_{\rm bol,cold}=L_{\rm BH,cold} = \epsilon_{0}\,\dot{M}_{\rm BH,cold}\,c^{2}\, \rm erg\,s^{-1};
\end{equation}
here the radiative efficiency $\epsilon_{0}=0.1$ and $\dot{M}_{\rm BH,cold}$ is the mass accretion rate
at the black hole horizon. In our model, we assume the radiation for both cold and hot modes
are isotropic and ISM is optically thin when we calculate the radiative transfer. The dust is
ignored in our model although it may dominate the radiation pressure in some regions of the galaxy.
The Compton heating and cooling are calculated in terms of ``Compton Temperature'' $T_{C}$, which
represents the energy-weighted average energy of photons emitted by the AGN. For the cold mode,
$T_{C,\rm cold} = 2\times 10^7$ K \citep{Sazonov:04}.

The character of the study of wind in  the hot mode is quite different compared to the cold
mode. It is much more difficult to detect wind in the hot mode. One reason is the dimness of
the sources hosting hot accretion flows. Another reason is that the temperature of wind in the hot
mode is too high thus the wind is usually fully ionized; it is therefore difficult to produce any
line features. Consequently, although we are gradually accumulating more and more observational
evidences for the existence of wind in hot accretion flows \citep[e.g.,][]{Wang:13, Tombesi:14,
Homan:16, Cheung:16, Ma:18}, the observational data is so far not good enough to  constrain the
properties of wind as in the case of cold mode. For example, in \citet{Wang:13}, by modeling
the iron emission line profile, we obtain the radial density profile of the hot accretion flow
around the supermassive black hole in our Galactic center, from which we infer the existence of
wind. In another example, \citet{Cheung:16} detect bisymmetric emission line structure in the
polar direction of an low-luminosity AGN from which they can measure the spatial distribution
of velocity field of the emission clouds. They infer the existence of wind by explaining such
a velocity field.

The relevant study starts from the pioneer hydrodynamical simulation work by \citet{Stone:99}
and followed by both analytical works and hydrodynamical and magneto-hydrodynamical numerical
simulations \citep[e.g.,][]{Yuan:12b, Yuan:12a, Narayan:12, Li:13, Sadowski:13, Yuan:15,
Bu:16a, Bu:16b, Amin:16}.  Different from the wind from cold mode,  theoretical study of wind
from hot mode is much easier. On one hand, radiation can be neglected for wind study in the
hot mode. On the other hand, it is much easier to simulate a geometrically thick accretion
flow than a thin disk. In spite of these, the study of the wind properties  in the hot mode
is not trivial, because one has to face with the challenge of how to discriminate winds (i.e.,
``real outflow'') from turbulence (see discussions in \citealt{Stone:99}).

By applying the ``virtual particle trajectory'' approach and based on  three dimensional GRMHD
simulation, \citet{Yuan:15} successfully  overcome this difficulty and calculate the wind
properties, including mass flux, velocity, and spatial distribution:
\begin{equation}
    \dot{M}_{\rm W,H} \approx \dot{M}(R_{\rm in})\left[ 1 - \left( \frac{3\,r_{s}}{r_{\rm tr}} \right)^{0.5} \right],
\end{equation}
\begin{equation}
    v_{\rm W,H}\approx (0.2-0.4)\,v_{\rm K}(r_{tr}),
\end{equation}
where $r_{s}\equiv 2GM_{\rm BH}/c^2$ is the Schwarzschild radius, $v_{\rm K}$ is the Keplerian
velocity, $r_{\rm tr}$ is the truncation radius of the outer thin disk described by \citet[see][and
references therein]{Yuan:14},
\begin{equation}\label{eq:truncR}
        r_{\rm tr} \approx 3\,r_{s} \left[ \frac{2\times10^{-2}\,\dot{M}_{\rm Edd}}{\dot{M}(R_{\rm in})} \right]^{2}.
\end{equation}
In the present work, we will adopt these theoretical results, because of our lack of good
observational constraints. We emphasize that the above results only apply to the case of ``SANE''
(i.e., standard and normal evolution) and a non-spinning black hole. It is needed to  calculate
the cases of ``MAD'' (i.e., magnetically arrested disk) and a spinning black hole (Yang, Yuan
et al. 2019, in preparation).

As for the angular distribution of wind from hot accretion flow, based on the detailed analysis
presented in \citet{Yuan:15}, we set that the mass flux of the wind is distributed mainly within
$\theta\sim 30^{\circ}$-$70^{\circ}$ and $110^{\circ}$-$150^{\circ}$ above and below the equatorial
plane, respectively. We note that the formation of Fermi bubbles detected in the Milky Way galaxy
has been successfully explained by the interaction between wind launched from the hot accretion flow
around Sgr A* and the interstellar medium in the Galaxy \citep{Mou:14,Mou:15}.

We now calculate the radiation in the hot mode. The black hole accretion rate in the hot mode is
\begin{equation}
   \dot{M}_{\rm BH,hot} \approx \dot{M}(R_{\rm in}) \left( \frac{3\,r_{s}}{r_{\rm tr}} \right)^{0.5}.
\end{equation}
The radiative efficiency of the hot accretion flow is no longer a constant but a function of
accretion rate. It  is described by the following formula \citep{Xie:12},
\begin{equation}
    \epsilon_{\rm EM,hot}(\dot{M}_{\rm BH}) = \epsilon_{0} \left( \frac{\dot{M}_{\rm BH}} {0.1\,L_{\rm Edd}/c^{2}}\right)^{a},
\end{equation}
the values of $\epsilon_0$ and $a$ are given in \citet{Xie:12} and we copy them here:
\begin{eqnarray}
  (\epsilon_0, a) &=& \left\{
    \begin{array}{ll} (0.2,0.59), & \dot{M}_{\rm BH}/\dot{M}_{\rm Edd}\la 9.4\times 10^{-5} \\
  (0.045,0.27), & 9.4\times 10^{-5} \lesssim \dot{M}_{\rm BH}/\dot{M}_{\rm Edd} \lesssim 5\times 10^{-3} \\
  (0.88,4.53), & 5\times 10^{-3}\lesssim \dot{M}_{\rm BH}/\dot{M}_{\rm Edd} \lesssim 6.6\times 10^{-3} \\
  (0.1,0), & 6.6\times 10^{-3}\lesssim \dot{M}_{\rm BH}/\dot{M}_{\rm Edd} \lesssim 2\times 10^{-2}
    \end{array} \right.
  \label{efficiencyfit}
\end{eqnarray}
The comparison of the radiative efficiency between the cold and hot modes has been shown in the
right panel of Figure~\ref{fig:anal}. Note that the efficiency of the hot mode is comparable to that
of the cold mode when the accretion rate is high. The radiation emitted from a hot accretion flow
has relatively more hard photons compared to that from a thin disk; thus, the Compton temperature of
a hot accretion flow is higher, $T_{\rm C,hot}\sim 10^8 {\rm K}$ \citep{Xie:17}.  Such a high
$T_{\rm C,hot}$ results in a relatively efficient radiative heating and is likely the reason for the
importance of radiative feedback in the hot mode, as we will discuss in \S\,\ref{subsec:hotfb}.

We note that in the current work we ignore dust in the ISM when we calculate the radiative
feedback.  If the dust were to be included, the opacity could be orders of magnitude larger
than the electron-scattering opacity \citep[e.g.,][]{Novak:12}; thus, a much larger portion of
radiation could be deposited in the ISM. In this case, as shown by the right plot of Fig. 1 in
\citetalias{Yuan:18b}, if the AGN accretion rate is not very low, the momentum flux of radiation
will be comparable to or even larger than that of the wind. So radiative feedback will be more
important than what we show here. 

\subsection{Models}
\label{subsec:models}

To investigate the role of the hot mode feedback, in this paper we have produced four models, as
given in  Table~\ref{tab:model}.   In the model ``fullcoldFB'', we assume that the AGN feedback
occurs only in cold mode over the entire range of accretion rates.  In the model ``nohotFB'', when
the accretion rate is lower than the ``boundary'' between the cold and hot  modes, instead of
adopting the hot mode as in the fullFB model, we simply turn off the AGN, i.e., there will be no
radiation and wind at all in this case.  We compare the simulation results of these two models with
those of the ``fullFB'' model presented in \citetalias{Yuan:18b} (but with some improvements as
described in \S\,\ref{subsec:fullFB}).  To understand the respective role of radiation and wind in
the hot mode, we carry out two additional models without wind (mechanical) feedback (i.e.,
``nohotmechFB'') and radiative feedback (i.e., ``nohotradFB'') respectively.

The left panel of Figure~\ref{fig:anal} compares the radiation and wind power between the fullFB and
fullcoldFB models.  The differences are of course only in the regime of $L\la2\times 10^{-2}L_{\rm
Edd}$ and they are very significant.  Radiation in the fullcoldFB model is almost always stronger
than that in the fullFB model. Wind power in fullcoldFB model is also much stronger than that in the
fullFB model unless when the luminosity is lower than $\sim 6\times 10^{-4}L_{\rm Edd}$. As we will
show in \S\,\ref{sec:results}, these differences  produce important effects on the AGN feedback.

\begin{deluxetable}{ccc}[!htbp]
    \tablecolumns{3}\tabletypesize{\scriptsize}\tablewidth{0pt}
    \tablecaption{Description of the models \label{tab:model}}
    \tablehead{ \colhead{model} & \colhead{AGN$_{\rm cold}$} & \colhead{AGN$_{\rm hot}$}}
    \startdata
        fullFB   & yes & yes \\
        fullcoldFB & yes & yes$^{1}$ \\
        nohotFB  & yes &  no \\
        nohotradFB & yes & only AGN$_{\rm hot, mech}$ \\ 
        nohotmechFB & yes & only AGN$_{\rm hot, rad}$ \\ 
    \enddata
    \tablenotetext{1}{In the fullcoldFB model, the radiation and wind outputs of AGN always follows the prescriptions in the cold mode over the entire range of accretion rate. }
\end{deluxetable}

 \section{Results}\label{sec:results}

\subsection{Some results of the updated fullFB model}
\label{subsec:fullFB}

\begin{figure}[!htbp]
    \centering
    \includegraphics[width=0.5\textwidth]{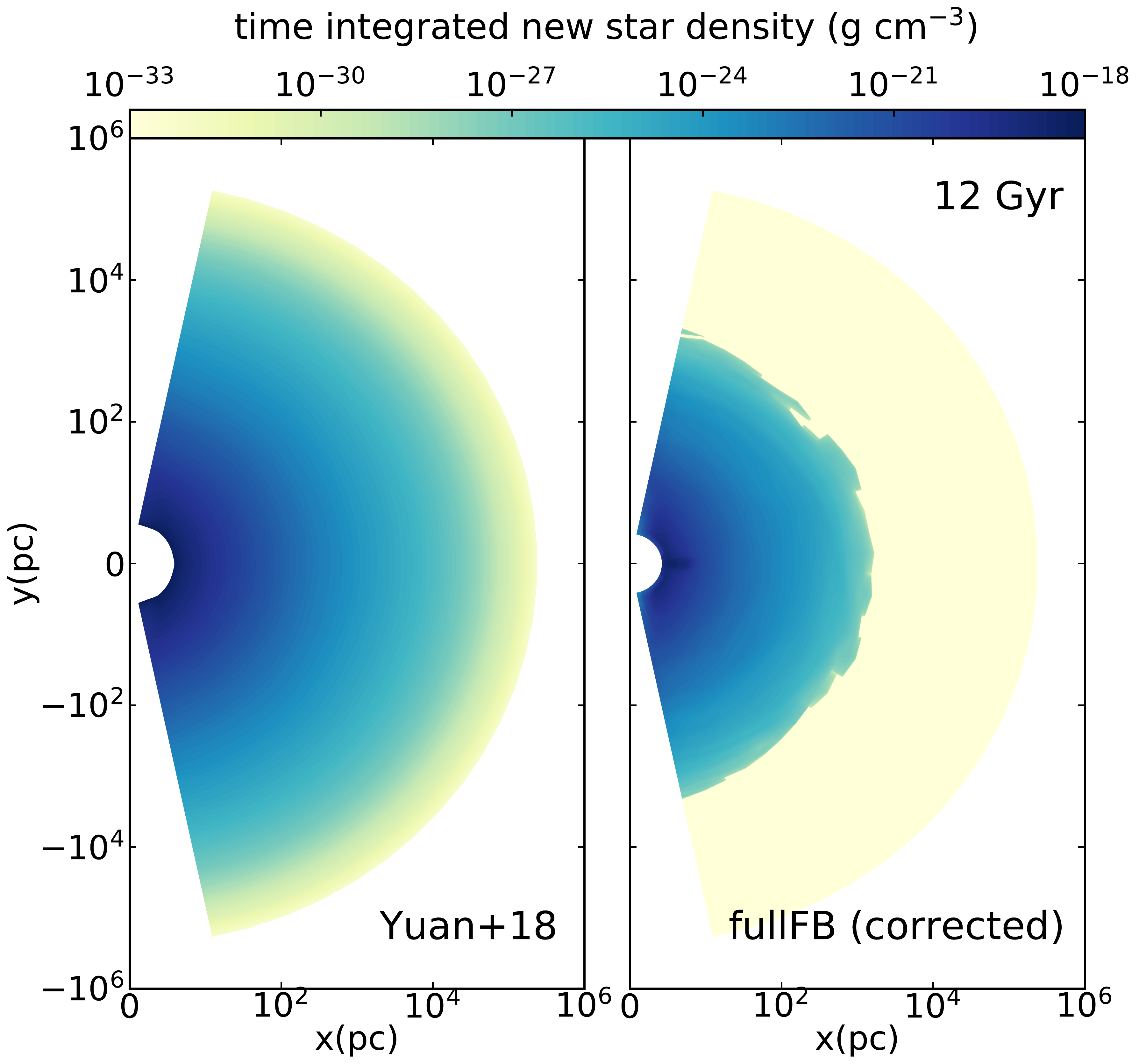}
    \caption{The mass density of the newly born stars, which are integrated for the entire simulation
    time. Left plot is the result from the fiducial model in \citetalias{Yuan:18b}, and the right
    plot is the ``corrected'' result from the present work, at which stars form only when the temperature is below
	$4\times10^4$ K and the { gas density is higher than $1 \rm \,cm^{-3}$.} }
    \label{fig:dnewst}
\end{figure}

For the fullFB model, we employ the AGN feedback for both cold and hot modes as mentioned above.
This model is identical to that in \citetalias{Yuan:18b} but with two improvements in the
present paper. One is on the calculation of star formation, as described in \S2.3.  The effect
of adopting the new star formation calculation is shown by Figure~\ref{fig:dnewst}.  We can see
that the amount of newly formed stars is significantly reduced compared to \citetalias{Yuan:18b},
and the distribution of the formed stars is highly concentrated within $\sim$ 1 kpc.  This is
in a good agreement with the recent observation \citep{Tadaki:18}, which showed that a large
fraction of stars forms in the central 1 kpc.

In addition to this improvement, we have also corrected a bug in computing the energy flux of the
wind for the fullFB model in \citetalias{Yuan:18b}. Although this does not change the overall
evolution of the black hole and the galaxy, we do find that it had over-produced the wind power. And
last, we would like to emphasize that, similar to \citetalias{Yuan:18b}, all models in this work
assume only a small degree of galactic rotation; thus there is no density-enhanced disk
\citep[see][as is the result for the model with higher degree of galactic rotation]{Yoon:18}.

\begin{figure}[!htbp]
    \centering
    \includegraphics[width=0.5\textwidth]{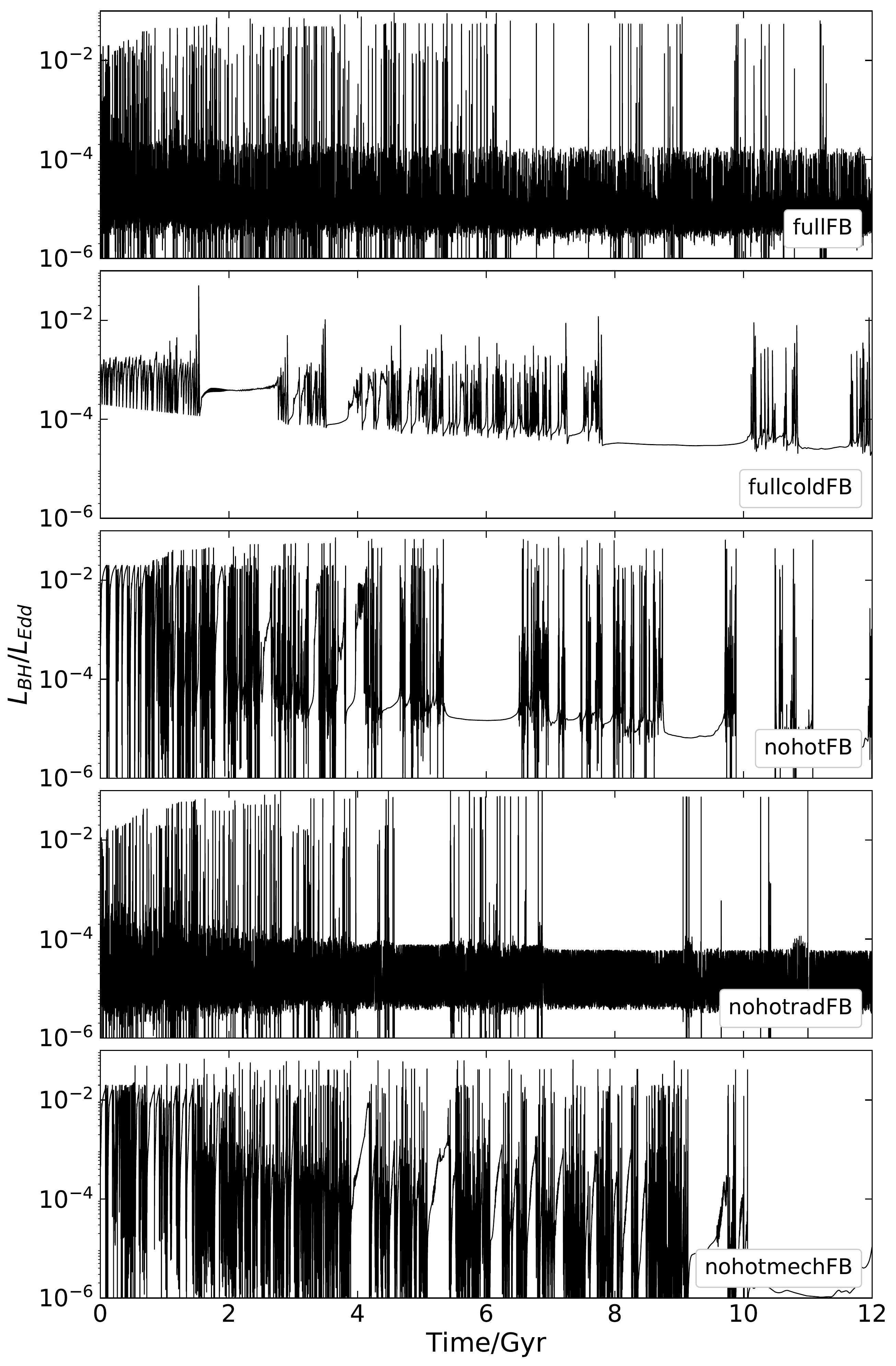}
    \caption{Light curves of AGN luminosity as a function of time for various models. }
    \label{fig:lightcurves}
\end{figure}

\begin{figure}[!htbp]
    \centering
    \includegraphics[width=0.5\textwidth]{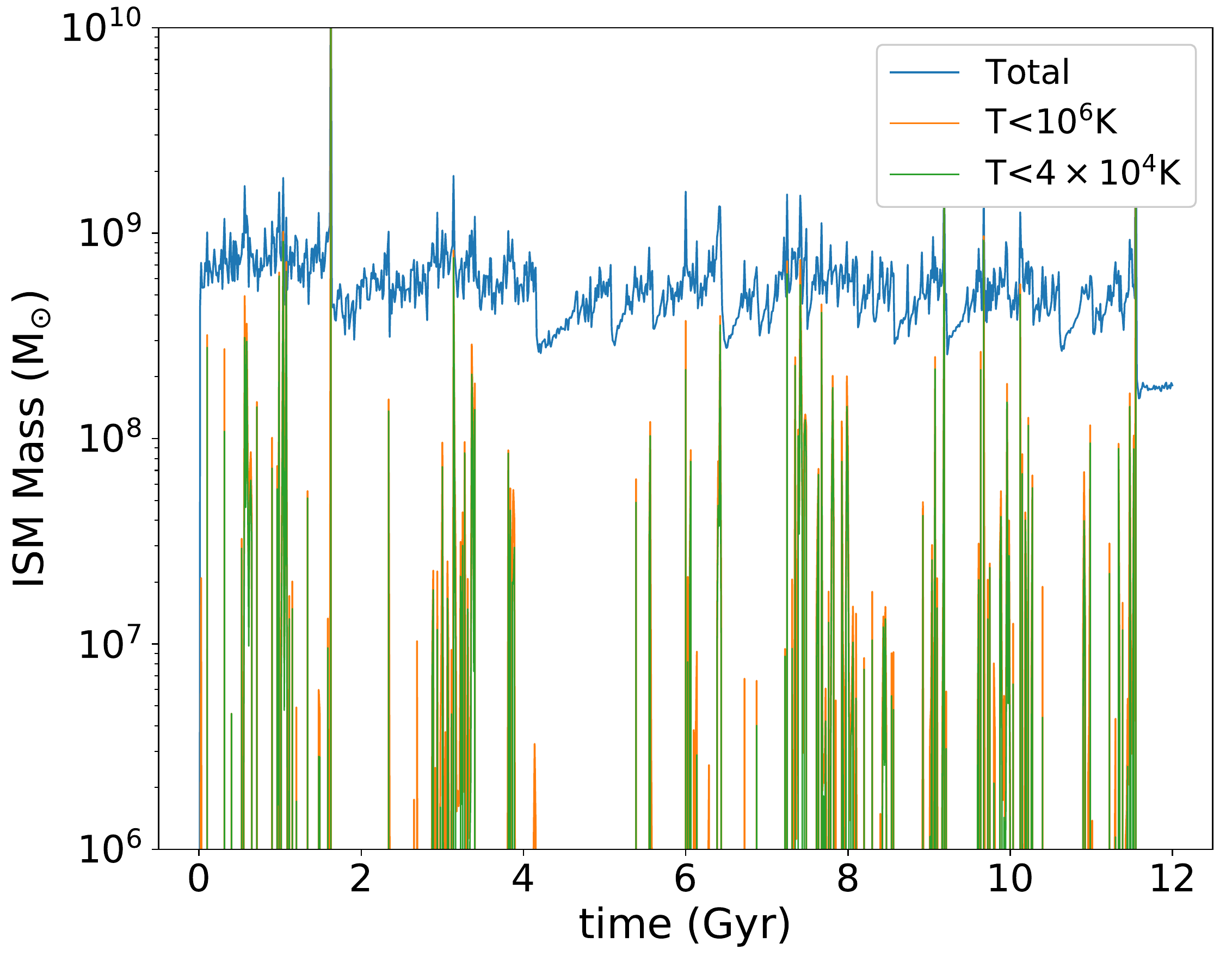}
    \caption{Mass budget of cold and hot ISM in the fullFB model.}
    \label{fig:ISM_full}
\end{figure}

Figure~\ref{fig:lightcurves} shows the AGN light curves. We can see that the overall shape is the same
as  in \citetalias{Yuan:18b}: it stays at low luminosity ($L_{\rm BH}\sim 10^{-5} L_{\rm Edd}$) for
most of lifetime with occasional bursts. These bursts occurs more frequently at earlier times as
a consequence of the more abundant gas supply.  In our galaxy configuration, the initial gas density in
the galaxy is very low and the stellar mass loss and SNe Ia are the main feedback for the
gas in the galaxy \citep{Pellegrini:12}.  Here, according to the stellar population synthesis model
that we adopt, most of stellar mass is lost at early times, resulting in more violent AGN feedback
at that stage.

Figure~\ref{fig:ISM_full} shows the total mass of the gas in the galaxy within different
temperature ranges.  While most of gas stays at $T>10^{7}$ K, it occasionally becomes cold enough for
star formation ($T<4\times10^{4}$ K) as a consequence of thermal instability
\citep[e.g.,][]{Li:14,Li:15}.


\begin{figure}[!htbp]
    \centering
    \includegraphics[width=0.5\textwidth]{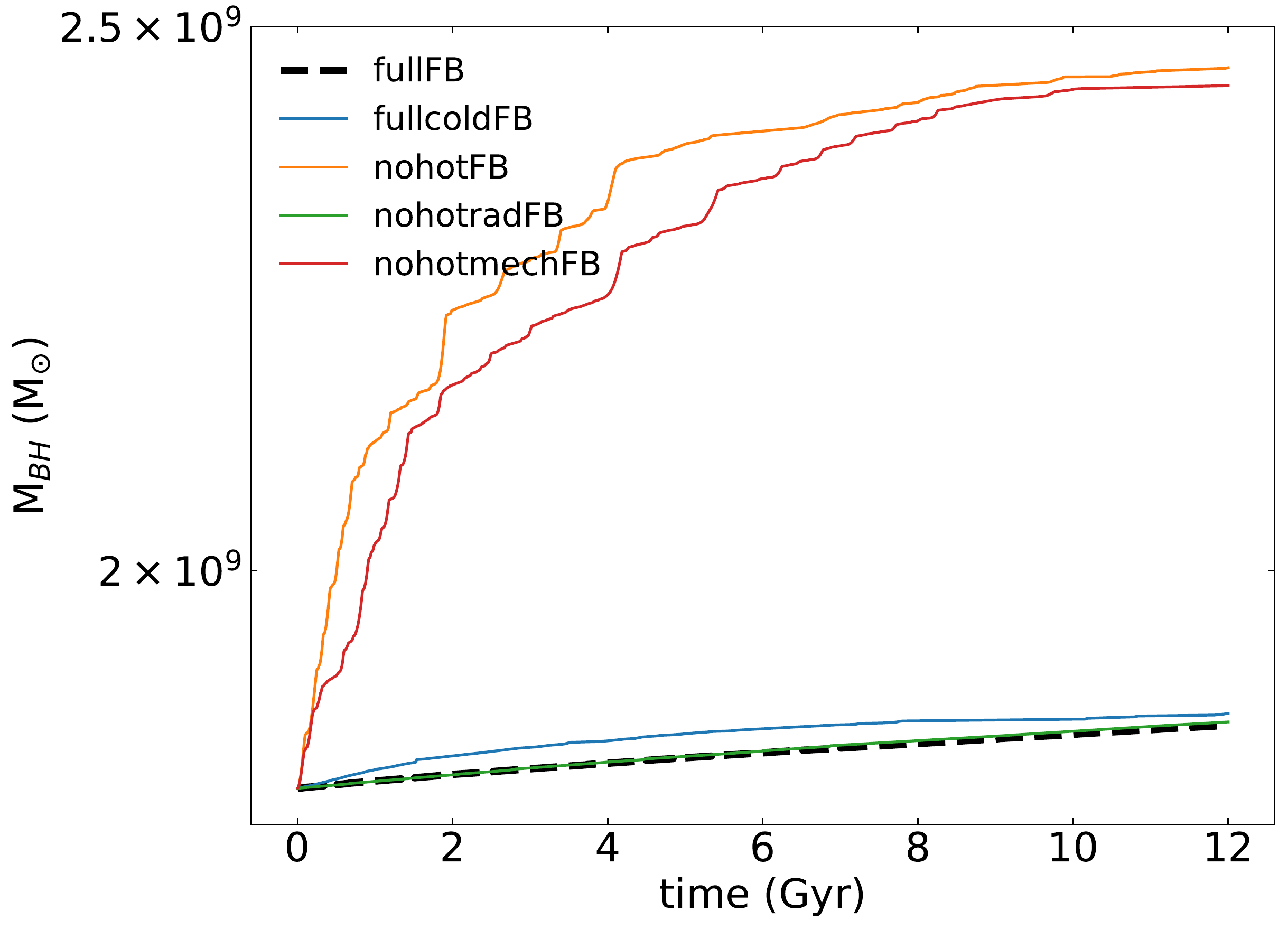}
    \caption{The evolution of black hole mass as a function of time for various models. }
    \label{fig:mbh}
\end{figure}

\begin{figure*}[!htbp]
    \begin{center}$
        \begin{array}{cc}
            \includegraphics[width=0.45\textwidth]{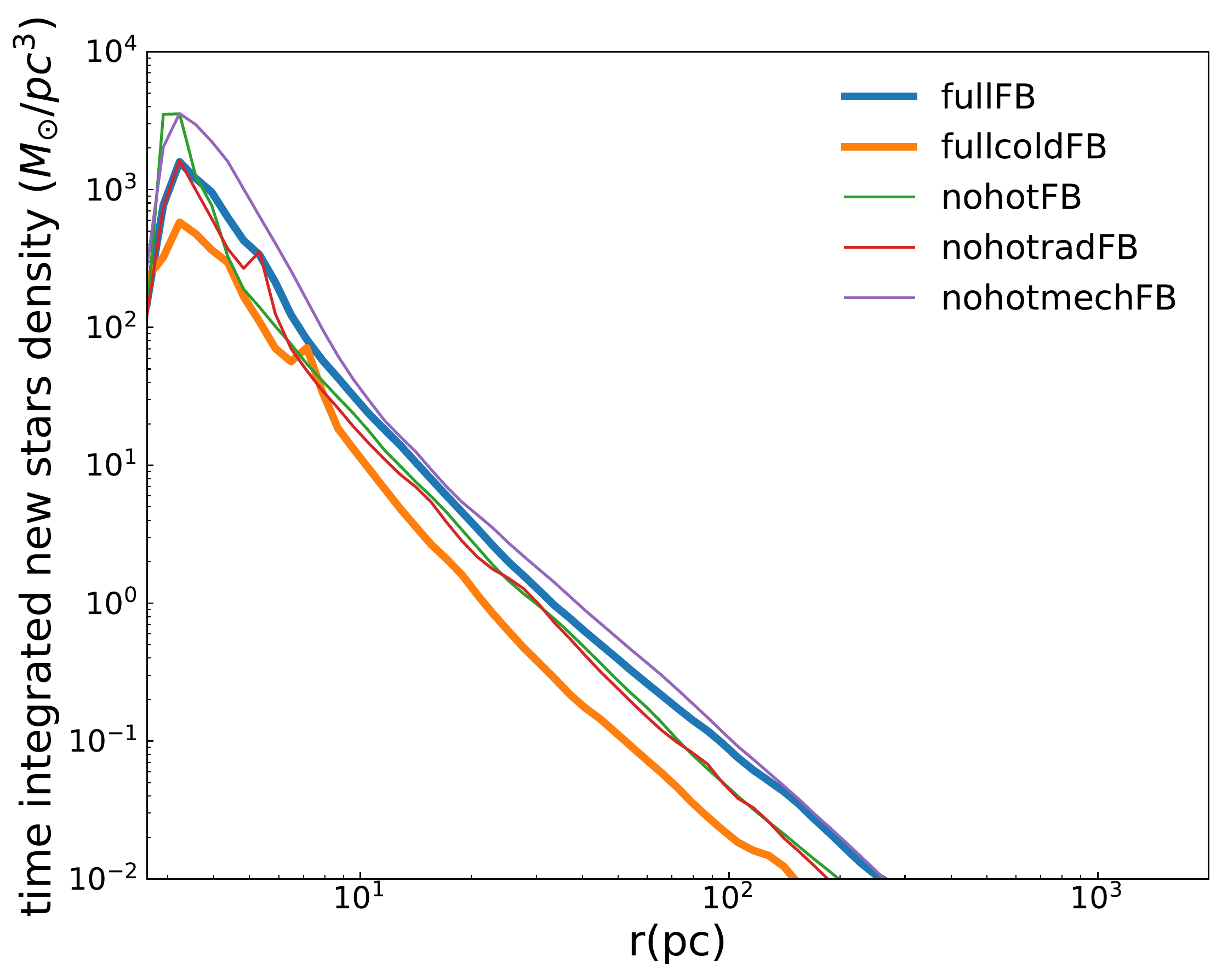} &
            \includegraphics[width=0.45\textwidth]{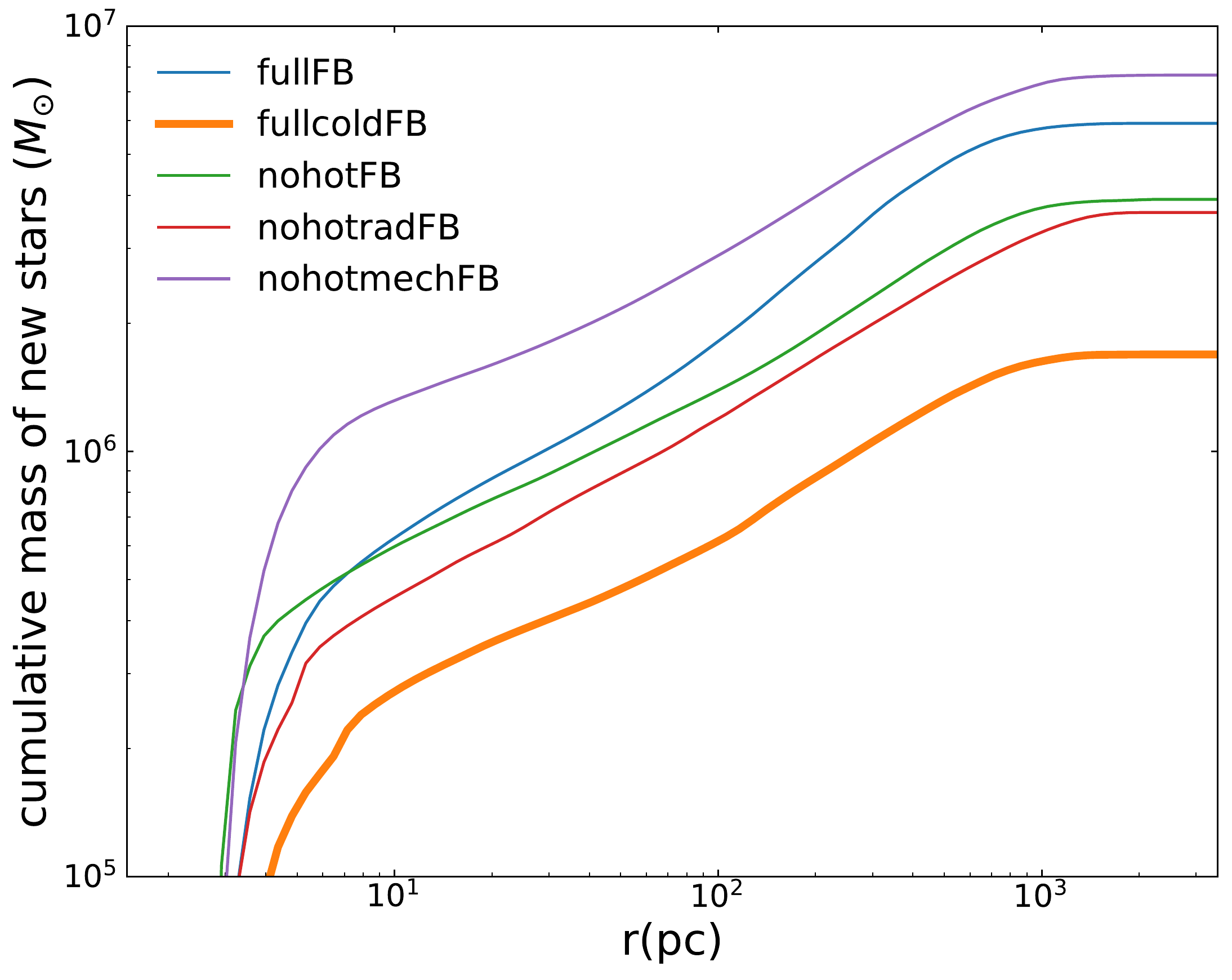}
        \end{array}$
    \end{center}
    \caption{ Left: The total mass of newly born stars at a given radius per unit volume.
              Right: The cumulative mass of the newly born stars.}
    \label{fig:newst}
\end{figure*}

\subsection{Comparison between fullFB and fullcoldFB models}

The resultant light curve for the fullcoldFB model is shown in Figure \ref{fig:lightcurves}.
Compared to the light curve of the fullFB model, we can see several obvious differences. One is that
the luminosity of the AGN is always below $10^{-2}L_{\rm Edd}$ in the fullcoldFB model while it can
be above this value in the fullFB model. The second difference is that we can clearly see a trend of
decreasing luminosity with cosmological evolution time in the fullcoldFB model while it is not so
obvious in the fullFB model. The  third difference is that the light curve in the fullcoldFB model is
much less bursty compared to the fullFB model.

\citetalias{Yuan:18b} have shown that the AGN accretion rate and light curve are  mainly controlled by
the wind feedback rather than radiative feedback. So we believe that the former two differences are
because of the stronger wind in the fullcoldFB model\footnote{Although the wind is stronger when
$L\la 6\times 10^{-4}L_{\rm Edd}$, both the wind mass flux and velocity are very small in that
regime (refer to eqs. \ref{coldwindflux} \& \ref{coldwindvelocity}); thus its role is not significant
compared to the case of $L\ga 6\times 10^{-4}L_{\rm Edd}$.}. When the wind becomes stronger, more
gas surrounding the black hole will be pushed away and thus the accretion rate will in general become
smaller. In addition, more gas will be blown out of the galaxy in the form of galactic wind; thus the
available gas for fueling the black hole will gradually become less. For the third difference
between the fullcoldFB and fullFB models mentioned above, we speculate that this is because of the
stronger radiation in the fullcoldFB model. The bursty feature of the AGN is likely because of the
accretion of small cold clumps by the black hole. The clumps are formed by thermal instability
\citep[e.g.,][]{McCourt:12, Sharma:12, Gaspari:13a, Li:14, Li:15, Qiu:18,Wang:19}. When radiation is
stronger, radiative heating becomes stronger; thus it is harder for the formation of clumps.

Figure~\ref{fig:mbh} shows the evolution of the black hole mass for each model. We can see that the
growth of black hole in both fullFB and fullcoldFB models is very little.

Figure~\ref{fig:newst} shows the radial mass distribution of newly born stars for various models,
which is integrated for the entire simulation time. The results of the fullFB and fullcoldFB
models are shown by the thick blue and thick orange lines respectively. It is apparent that star
formation is highly suppressed in the fullcoldFB model compared to the fullFB model.   The reason
for reduced star formation in fullcoldFB model may be twofold. On one hand, the AGN radiation is
stronger in the fullcoldFB model, resulting in stronger radiative heating to the gas in a large
region of the galaxy where the optical depth is smaller than unity \citepalias{Yuan:18b}. On
the other hand, the wind in the fullcoldFB model is also much stronger than in the fullFB
model. The strong wind blows away the gas in the galactic center up to distance of a few kpc,
which results in the decrease of density and suppression of star formation.  The star formation
perhaps can be enhanced temporarily at $\sim$ kpc scale, where the gas is compressed due to
the interaction between the wind and the ISM \citep{Cresci:15}.  However, such a temporal
enhancement is averaged out and not present in this time-integrated figure.

The right panel of Figure~\ref{fig:newst} shows the total mass of newly born stars for various
models. We can see that the total mass of new stars in the fullcoldFB model is an order of
magnitude smaller than in the fullFB model.

\begin{figure}
    \centering
    \includegraphics[width=0.5\textwidth]{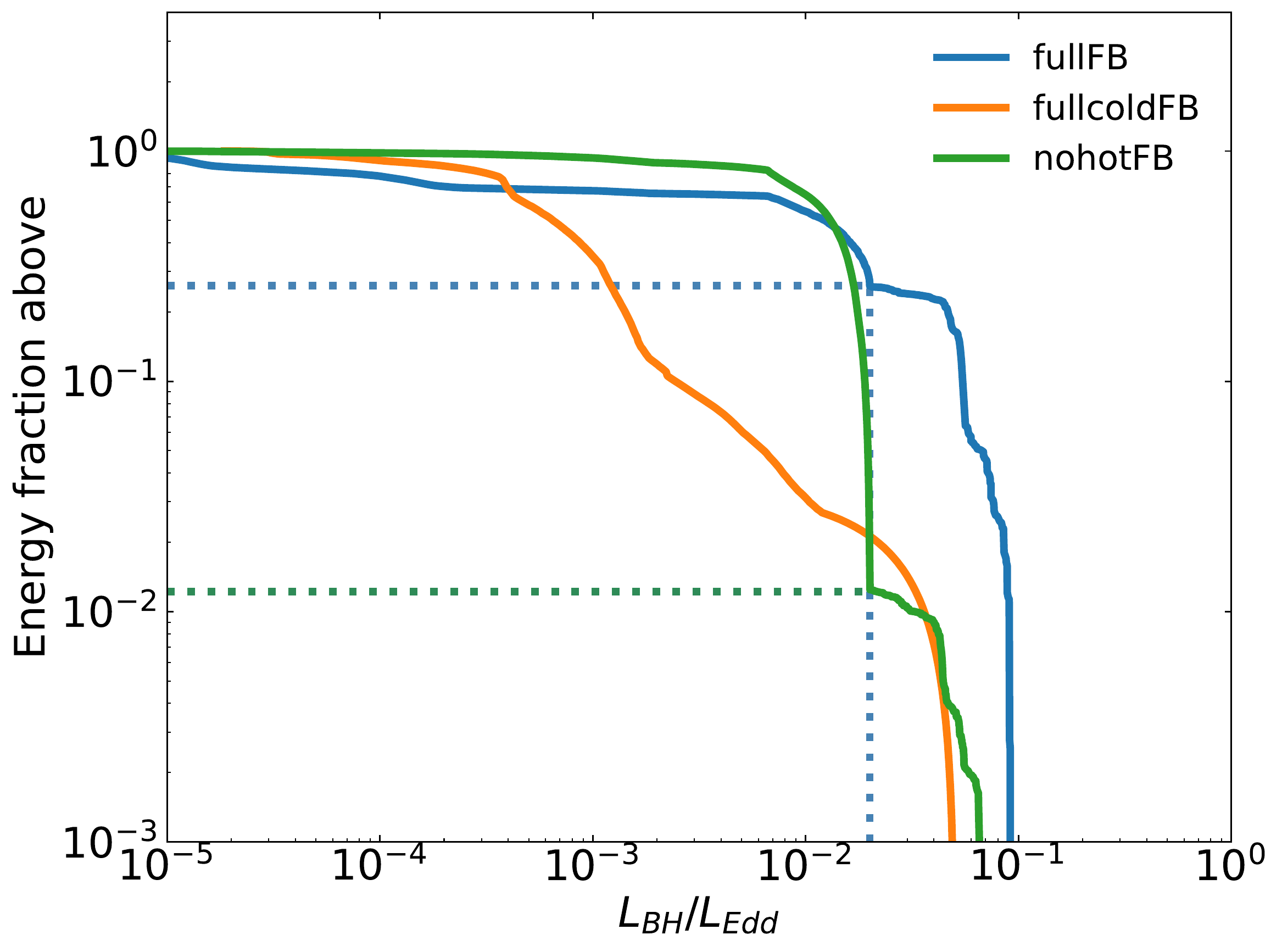}
    \caption{Percentage of the total cumulative energy emitted above the values of the
Eddington ratios. The horizontal dotted lines represent the portion of emitted energy above the Eddington ratio of 0.02.}
    \label{fig:duty_L}
\end{figure}

And last, let us examine the percentage of the emitted total energy above a given Eddington ratio.
Figure~\ref{fig:duty_L} shows the predicted results for various models. Observationally, it is
believed that AGNs spend most of their time in the low luminosity AGN phase but emit most of
their energy during the high-luminosity AGN phase \citep{Soltan:82, Yu:02}. For the fullFB
mode, the fraction of energy ejected above $2 \%L_{\rm Edd}$ is about 25\%\footnote{This value
is larger than that given in \citetalias{Yuan:18b},  which was only 6\%, and is more consistent
with the observational constraints. Such a difference is likely because of the two updates
of our model we have mentioned in \S\,\ref{subsec:fullFB}.}. But for the fullcoldFB model, 
the fraction becomes very small ($\sim$2\%), inconsistent with observations.

\subsection{Comparison between fullFB and nohotFB models}

The AGN light curve produced by the nohotFB model is shown in Figure~\ref{fig:lightcurves}. Unlike
the fullFB model, in which the AGN luminosity is  $\sim 10^{-5}\,L_{\rm Edd}$ for most of time,
the AGN luminosity in the nohotFB model becomes generally higher and fluctuates in the range
of $10^{-5} \la L_{\rm BH}/L_{\rm Edd}  \la 10^{-2}$. The higher ``average'' AGN luminosity
is obviously because of the absence of  AGN feedback when $L\la 2\%L_{\rm Edd}$. Due to the
absence of wind and radiation, the gas surrounding the black hole has averagely  higher density
and lower temperature and thus the accretion rate is higher.

In accordance with the AGN light curves, the black hole growth is also distinctly different
between the fullFB and nohotFB models, as shown by Figure~\ref{fig:mbh}. For the nohotFB model,
since the accretion rate is on average significantly higher than in the fullFB model, the black
hole can grow to a larger mass than in the fullFB model.  Although the hot mode occurs at low
accretion rates, this mode is not negligible because the time fraction of being in hot mode is
very large. This result strongly indicates the importance of including the hot mode feedback.

The radial distribution of newly born stars for the nohotFB model is shown in
Figure~\ref{fig:newst} by the thin green line.	Compared to the fullFB model, we can see that
in the nohotFB model, the star formation is enhanced within $\sim 4 {\rm pc}$, but reduced
outside of this radius.  In other words, the presence of the wind and radiation in the hot
mode suppresses  the star formation only in the vicinity of the black hole, which is apparently
surprising. Our explanation is as follows. The mechanical feedback by wind (maybe radiation also)
pushes gas outwards, affecting the star formation activity in a complicated way. On one hand
the wind can reduce star formation as it depletes the inner region in the galaxy. On the other
hand, it can also enhance star formation as it can make the gas inhomogeneous by compressing
the gas. The net effect on the total star formation may depend on the power of the wind. If
the wind is very powerful, it can push a lot of gas to large radius where star is difficult to
form since the density there is too low. If the wind is weak, it cannot push the gas too far
away and its main role is to make the gas inhomogeneous thus helping the star formation. Since
the wind in the hot mode is somewhat weak, it suppresses the formation only in the small radii
(i.e., $\le 4 {\rm pc}$) where it is still energetic enough to push the gas away.  At large radii
(i.e., $\ge 4 {\rm pc}$), the gas is just perturbed by the wind but not pushed away, thus star
formation is enhanced.	As we can see in the right panel of Figure~\ref{fig:newst}, the total
mass of newly born stars in the nohotFB model is smaller than in the fullFB model. We note that
the wind in the cold mode is powerful enough to push the gas to large radius where star formation
is hard since the density is too low, so wind in the cold mode always suppresses star formation.

The percentage of the total energy emitted above a given value of Eddington ratio for the nohotFB
model is shown in Figure~\ref{fig:duty_L}. We can see that, similar to the case of fullcoldFB
model, the percentage of energy emitted above $2\%L_{\rm Edd}$ for nohotFB model is very small,
$\sim 1\%$, which is inconsistent with the observations. This again indicates  that the hot
feedback mode is important.

\subsection{The roles of radiation and wind in the hot feedback mode}\label{subsec:hotfb}

We have also run two additional models to study the roles of radiation and wind in the hot
mode, i.e., the nohotradFB model and nohotmechFB model. We find that the overall results of
the nohotradFB  model are mostly the same as the fullFB model. This implies that, in the low
accretion regime, the mechanical feedback is likely dominant over the radiative feedback.

For the nohotmechFB model, we find that the AGN luminosity light curve and star formation activity
have similar shapes with the nohotFB model, which can also be explained by the dominance of
mechanical feedback in the hot mode. However, we find that the black hole growth is more suppressed
in the nohotmechFB model compared to the nohotFB model. This is likely because radiation in the hot
feedback mode can heat the gas surrounding the black hole thus reducing the mass accretion rate of the
black hole. This indicates that, depending on the physical questions of interest, radiation in the
hot mode still can play an important role and cannot be  neglected.

\section{Summary and conclusion}\label{sec:summary}

Recently we have continued our study of the effects of AGN feedback in an isolated elliptical
galaxy \citep{Yuan:18b,Yoon:18, Li:18, Gan:19}, along the line of our previous works
\citep[e.g.,][]{Ciotti:01, Ciotti:07, Ciotti:10, Gan:14}. The main improvement of this series
of works is the incorporation of  state-of-art physics of black hole accretion in the model,
including the exact discrimination of the hot and cold accretion modes and,  more importantly,
the recent important progresses in our understanding of radiation and wind  in the hot accretion
mode. The adopted accretion physics has been presented in detail in \citetalias{Yuan:18b}
\citep[see also][]{Yuan:18a}. These works have focused on different aspects of the problem:
low-specific angular momentum  galaxy \citepalias{Yuan:18b}, high-specific angular momentum
galaxy \citep{Yoon:18}, the different roles of AGN and stellar feedback \citep{Li:18}, and the
role of gravitational instability of the gaseous disk \citep{Gan:19}.

While consensus has been reached that the accretion rate of the central black hole covers a wide
range and the AGN in most galaxies pass through both cold and hot modes during their evolution,
most previous work in this field either focuses only on one mode; or even if they include both,
the accretion physics is not correctly adopted.  The aim of the present work is to study how
important it is to correctly include both modes. For this aim, we specifically focus on the
wind and radiation feedback in the hot mode (but the jet is neglected in this work).

We have run two test models, namely the fullcoldFB model and the nohotFB model (see
Table~\ref{tab:model}). In the fullcoldFB model,  the AGN always stays in the  cold mode,
no matter what value the accretion rate is. In the nohotFB model, we simply turn off the AGN
feedback once the accretion rate is smaller than $0.2L_{\rm Edd}/c^2$ (i.e., the AGN enters
into the hot mode).  We then compare the simulation results, such as AGN light curve and  star
formation, of these two models with the fullFB model in which  both modes are correctly included.

For the fullcoldFB model, the wind and radiation outputs from the AGN are in general much stronger
than in the fullFB model (the left panel of Figure~\ref{fig:anal}). These strong winds push the
gas away from the black hole  and even out of the galaxy. Consequently, the luminosity of the AGN
becomes significantly lower and decreases with time (Figure~\ref{fig:lightcurves}). Since the
AGN rarely stays in the high-luminosity regime in the fullcoldFB model,  the percentage of the
emitted total energy in the luminous regime becomes much smaller compared to the fullFB model,
inconsistent with observational constraints (Figure~\ref{fig:duty_L}). The stronger radiation
in the fullcoldFB model also makes the thermal instability of the gas in the host galaxy
harder to occur and thus fewer clumps will be formed and the AGN light curve is less bursty
compared to the fullFB model (Figure~\ref{fig:lightcurves}). The strong wind and radiation in
the fullcoldFB model also strongly suppress the star formation in the host galaxy, so the total
mass of newly born stars becomes an order of magnitude smaller compared to the fullFB model
(Figure~\ref{fig:newst}). This is because the strong winds blow the gas far away from the central
region of the galaxy and the strong radiation also heats the gas in a large region of the galaxy.

For the nohotFB model, the AGN luminosity averagely becomes significantly higher compared to the
fullFB model. This is because the density of the gas surrounding the black hole becomes higher and
temperature becomes lower  due to the absence of the wind and radiation when the AGN accretion
rate is below $0.2L_{\rm Edd}/c^2$. In accordance with this change, the black hole mass also
becomes  higher in the nohotFB model (Figure~\ref{fig:mbh}). It is interesting to note that,
compared to the fullFB model, in the nohotFB model  star formation is enhanced only at $r\le 4
{\rm pc}$, but is reduced in all the region beyond that radius. We speculate that the reason is
that the power of wind in the hot mode is weak. In this case, winds can only be able to  blow
the gas in the central region of the galaxy away. Beyond that radius, they mainly play a role
of disturbing the gas and making the gas inhomogeneous, which is helpful for star formation.

In addition to the fullcoldFB and nohotFB models, we have also run two additional models,
namely nohotradFB and nohotmechFB models (Table~\ref{tab:model}), to examine the respective
roles of wind and radiation in the hot feedback mode.  It is found that the overall results
of the nohotradFB model are similar to the fullFB model, which implies that wind rather than
radiation plays a dominant role in the hot feedback mode. This is further confirmed by the
comparison between nohotmechFB and nohotFB models, which shows that their AGN light curve and
star formation activity are similar.  However, we find that the black hole growth is more
suppressed in the nohotmechFB model compared to the nohotFB model, which is because of the
additional radiative heating in the nohotmechFB model causing a decrease of the accretion rate.
This indicates that, depending on the problem of our interest, radiation sometimes also plays
an important role.  Combined with \citetalias{Yuan:18b}, this paper has found that the wind
feedback dominates for controlling the black hole growth and the star formation activity,
but the effect of radiative feedback  is not negligible.

These results  indicate that the hot mode plays an important role in AGN feedback thus cannot
be neglected; it is crucial to correctly include both modes.

Some caveats exist in the present work and we plan to investigate them in the future: 1) we only
consider the low-angular momentum galaxy; 2) we have not taken into account the external gas supply
to the galaxy; 3) the jet is neglected; 4) dust has not been included. The quantitative results may
change after we take into account these effects but we expect that the major conclusions of the
present paper should remain unchanged.

\section*{Acknowledgments}
We are grateful to the referee for the constructive comments which have significantly improved
our paper. We thank Drs. Yuan Li and Miao Li for useful discussions.  DY and FY are supported in
part by the National Key Research and Development Program of China (Grant No. 2016YFA0400704),
the Natural Science Foundation of China (grants 11573051, 11633006, 11650110427, 11661161012),
the Key Research Program of Frontier Sciences of CAS (No.  QYZDJSSW-SYS008), and the Astronomical
Big Data Joint Research Center co-founded by the National Astronomical Observatories, Chinese
Academy of Sciences and the Alibaba Cloud.  This work made use of the High Performance Computing
Resource in the Core Facility for Advanced Research Computing at Shanghai Astronomical Observatory.

\bibliographystyle{aasjournal}
\bibliography{AGNHotFB}
\end{document}